\documentclass[reprint,amsmath,amssymb,aps,prl,nobalancelastpage]{revtex4-2}

\usepackage{graphicx}
\usepackage{dcolumn}
\usepackage{bm}
\usepackage{amsmath}
\usepackage{amssymb}
\usepackage{bm}
\usepackage{physics}
\usepackage{graphics}
\usepackage{graphicx}
\usepackage[numbib]{tocbibind}
\usepackage[bottom]{footmisc}
\usepackage{float}
\usepackage{verbatim}

\begin{document}

\title{Vacuum Wannier Functions for First-Principles Scattering and Photoemission}

\author{Tyler Wu}
\affiliation{Cornell University}
\author{Tomás Arias}
\affiliation{Cornell University}

\date{\today}

\begin{abstract}
We establish a first-principles theory of vacuum Wannier functions unifying tight-binding and nearly–free-electron descriptions across solid–vacuum interfaces. Analytic solutions for canonical Wannier functions in arbitrary dimension and disentangled functions in 1D motivate a numerically verified 3D Wannier close-packing principle, enabling dense k-space construction of full Born-series scattering states at interfaces and thus predictive photoemission calculations without semiempirical vacuum potentials. Applications to graphene and h-BN reveal corrections beyond the first-Born approximation.
\end{abstract}

\maketitle

Wannier functions have become indispensable in modern electronic structure theory. Their real-space locality enables efficient band-structure interpolation~\cite{RevModPhys.84.1419}, response-function calculations~\cite{PhysRevX.11.041053}, and linear-scaling many-body methods~\cite{PhysRevLett.87.246406}, while also providing a common framework for polarization, orbital magnetization, and topological classification~\cite{vanderbilt2018berry}. Beyond electronic-structure methods, Wannier functions have been applied across diverse fields, including superconductivity~\cite{PhysRevB.76.165108}, plasmonics~\cite{Sundararaman2014}, photonics~\cite{Leung:93,PhysRevB.88.075201}, and cold-atom lattice physics~\cite{PhysRevE.66.046608}.

Despite this broad applicability, the use of Wannier functions has remained largely confined to bulk material systems. Extending their utility beyond bulk systems is necessary to understand coupling at interfaces between condensed media and surrounding vacuum or dielectric environments. A unifying challenge thus arises in understanding phenomena such as photoemission~\cite{RevModPhys.93.025006}, electron scattering~\cite{PhysRevB.70.245322,Werner2023PartialIntensities}, field emission~\cite{PhysRevB.100.165421}, and optical reflection and tunneling~\cite{Noda2024PCSELReview}. In such processes, fields and excitations propagate from localized or periodic states within the medium into open continuum states outside it. Developing a material-specific first-principles framework for this vacuum–-bulk coupling remains a major challenge~\cite{PhysRevB.78.165406,PhysRevB.104.115132,PhysRevB.70.245322} and is rarely addressed in standard plane-wave packages owing to periodic-image effects.

As a representative example, consider photoemission, where electrons excited in the solid must propagate into the vacuum. Current treatments often rely on the three-step picture~\cite{10.1117/12.158575,PhysRevB.104.115132}, invoking phenomenological models of the emission process such as simple one-dimensional potential steps~\cite{PhysRevB.101.235447}. One-step approaches combining Wannier or tight-binding initial states with plane-wave final states~\cite{yen2024first,day2019chinook,PhysRevResearch.5.033075} provide greater formal unity but still neglect the material-specific Born-series structure of continuum states near the surface, where matrix elements exert the strongest influence. A natural resolution is the construction of regular lattices of maximally localized Wannier functions in vacuum, enabling arbitrary vacuum extension without the cost of correspondingly large DFT supercells. However, the theory of such vacuum Wannier functions has remained underdeveloped.

To establish the theoretical foundation for \emph{vacuum Wannier functions} (VWFs), we derive analytic solutions corresponding to the global minimizers of the Marzari--Vanderbilt localization functional in arbitrary dimension \(d\), and of the Souza--Marzari--Vanderbilt disentanglement procedure~\cite{RevModPhys.84.1419} in one dimension, complemented by numerical solutions of the disentanglement problem in three dimensions. These results yield a family of highly localized basis functions centered on regular close-packed lattices, ideally suited for arbitrary extension of the vacuum region in supercell calculations without the need for additional DFT computations.

To demonstrate the utility of this framework, we perform quantitative photoemission calculations for graphene and hexagonal boron nitride, including near-threshold emission observables such as the mean transverse energy (MTE), relevant to photocathode performance, and the rms longitudinal spread, relevant for ultrafast electron diffraction (UED). We find that photoemission from these systems depends sensitively on the interplay of symmetry, matrix elements, and higher-order Born effects, all of which are naturally captured within our framework.

\begin{figure}[t]
    \centering
    \includegraphics[width=\linewidth]{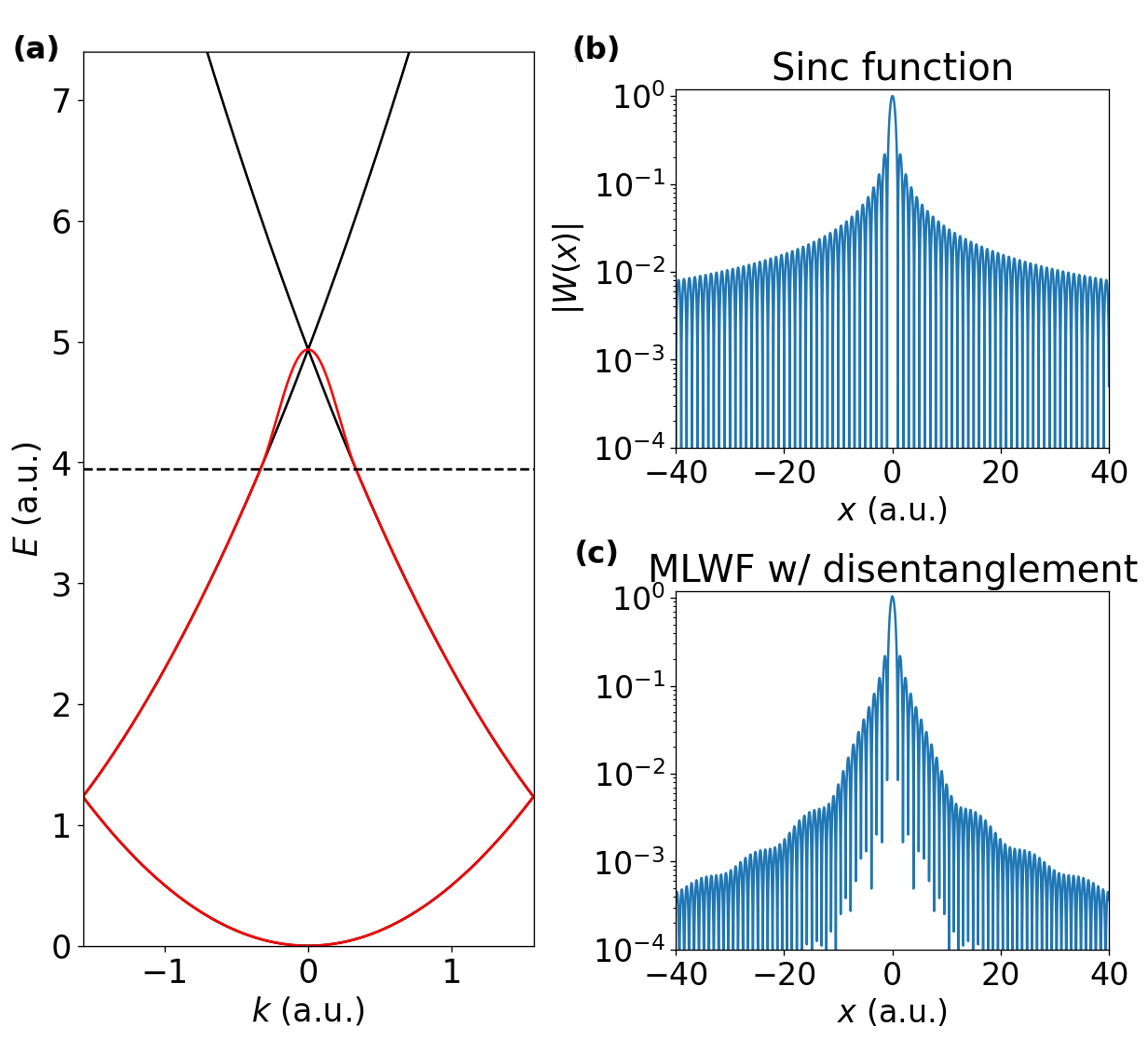}
    \caption{(a) Dispersions for an empty 1D lattice (black) and the two disentangled bands obtained from the analytic $\theta(k)$ interpolation (red). The dashed line marks the frozen window energy boundary. (b) and (c) compare the decay of the magnitudes of the sinc-like and analytically disentangled MLWFs, respectively. }
    \label{fig:1D bands}
\end{figure}

\emph{Challenges of Wannier localization in the vacuum limit.---}
Constructing Wannier functions (WFs) in systems containing large vacuum regions—such as slab geometries—presents subtle challenges. As the vacuum thickness increases, nearly free-electron bands proliferate, forming a quasi-continuum of delocalized states. In this regime, the standard Marzari–Vanderbilt (MV) localization procedure becomes problematic. The MV spread functional, $\Omega = \Omega_I + \tilde{\Omega}$, separates into a gauge-invariant term $\Omega_I$, set by the choice of band subspace, and a gauge-dependent term $\tilde{\Omega}$ minimized by unitary rotations at each $k$-point. In the vacuum limit, Bloch states reduce to plane waves, and the WFs acquire sinc-like envelopes with algebraic $1/x$ decay and divergent second moments (Fig.~\ref{fig:1D bands}(b)), rendering the functional ill-defined. Numerically, the Souza–Marzari–Vanderbilt (SMV) disentanglement procedure~\cite{doi:10.1137/18M1167164} can remove the formal divergence, but the resulting Wannier functions retain large residual spreads, which render the localization landscape ill-conditioned; this ill-conditioning, in turn, leads to instabilities and irregular Wannier-center positions that obstruct straightforward vacuum padding.

\emph{Canonical vacuum minimizers: regular lattice structure.---}
In the standard textbook single-band construction for the one-dimensional empty lattice dispersion relation (Fig.~\ref{fig:1D bands}(a)), the resulting Wannier functions are the familiar sinc functions (Fig.~\ref{fig:1D bands}(b)), centered on lattice points and exhibiting algebraic $1/x$ decay and divergent variance~\cite{PhysRevLett.86.5341}. Despite the divergence, this solution fully minimizes to zero the (finite) gauge-dependent part of the spread.

Using a unit-cell partitioning procedure, we have shown analytically that this result generalizes to multidimensional ($d>1$) and multiband ($N>1$) cases. This yields $N$ identical sinc-like Wannier functions centered on a regular grid within the unit cell that minimize to zero the gauge-dependent part of the spread functional. (See End Matter for the analytic demonstration of this absolute minimization.)

\emph{Disentanglement and optimal localization.---}
To cure the divergence of the gauge-invariant part of the spread functional identified above in vacuum, and to optimize the Wannier spread in the presence of overlapping bands, a standard approach is the Souza–Marzari–Vanderbilt (SMV) disentanglement procedure~\cite{doi:10.1137/18M1167164}. Disentanglement directly targets the gauge-invariant part of the spread,
\begin{equation}\label{eq:Omega_I}
\Omega_I=\frac{a}{4\pi}\int_{-\pi/a}^{\pi/a} \! dk\;
\mathrm{Tr}\!\bigl(\partial_k P(k)\,\partial_k P(k)\bigr),
\end{equation}
where \(P(k)\) is the projector onto the periodic Bloch states used to construct the Wannier functions~\cite{RevModPhys.84.1419}. In practice, a \emph{frozen-window} constraint is imposed to exactly reproduce a chosen set of bands within a fixed energy interval. (Fig.~\ref{fig:1D bands}(a) illustrates the procedure for a simple $N=2$ example.) Operationally, this is enforced by starting from the exact eigenstates of that subspace and augmenting with higher-energy states to improve $\Omega_I$. Specifically, in regions where the frozen window contains fewer bands than the target number of Wannier functions, the algorithm supplements the basis with smoothly varying unitary mixtures drawn from an outer window, which must extend sufficiently far in energy to fully capture the bands that exit the inner window. The result is a subspace of consistent dimension and a globally smooth projector suitable for constructing localized Wannier functions.

Prior work on disentanglement of vacuum states was limited to numerical tests in $d=1,2$ dimensions. In pioneering work, Damle, Levitt, and Lin~\cite{doi:10.1137/18M1167164} observed that disentanglement regularizes \(\Omega_I\), producing regularly spaced Wannier functions with stronger algebraic decay (\(\sim x^{-2}\)) and finite variance. They further found evidence that when the gauge is made absolutely smooth (\(C^\infty\)), the decay becomes super-algebraic; however, they were unable to construct functions that are orthogonal or reproduce the exact inner eigenspace. They neither establish that the resulting regularly spaced functions globally minimize the gauge-dependent spread functional $\tilde{\Omega}$ nor explore beyond \(d=2\) dimensions.

To extend these results, we have developed a fully analytic solution (Fig.~\ref{fig:1D bands}(c)) to the SMV disentanglement procedure for vacuum in $d=1$ (see End Matter). This solution renders the disentanglement transparent, globally minimizes $\tilde{\Omega}$, and again yields Wannier centers at regular spacing. Because real applications require $d=3$ dimensions, we consider this case next.

\emph{Close-packed lattice ansatz in $d$ dimensions.---}
Guided by our exact analytic results for the canonical case in $d=3$ and the disentangled case in $d=1$, and inspired by the two-dimensional numerical experiments of Damle, Levitt, and Lin, we performed three-dimensional numerical calculations for Wannier functions initialized on cubic lattices. We found that the resulting maximally localized functions obtained via disentanglement distort their centers to form locally close-packed structures. This motivates a conjecture for arbitrary dimensions $d$: \emph{given an appropriate outer window, disentangling and optimally Wannierizing $N$ bands in the empty lattice yields symmetry-equivalent MLWFs on a regular close-packed lattice}.

\begin{figure*}[ht]
    \centering
    \includegraphics[width=0.85\linewidth]{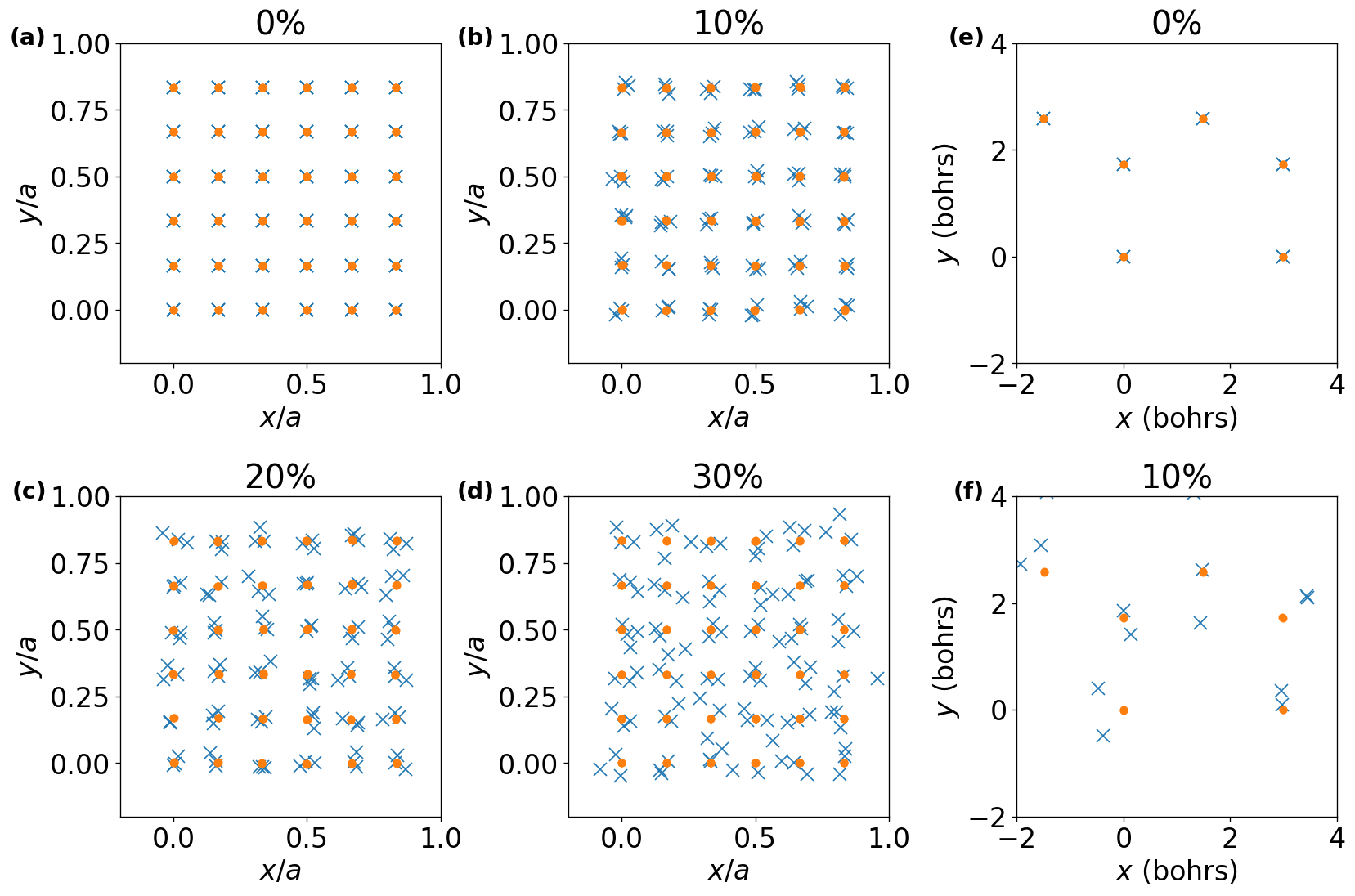}
    \caption{Stability tests of close-packed lattices for MLWF centers in vacuum: (a–d) projected view of fcc arrangement in a cubic supercell; (e–f) projected view of hcp arrangement in a hexagonal supercell. Blue $\times$'s mark initial randomized centers, and orange circles mark the relaxed positions. Percentages denote rms displacements (relative to the nearest-neighbor spacing) of the initial Gaussian-distributed displacements.} 
    \label{fig:perturbed_centers}
\end{figure*}

Motivated by the fact that the face-centered cubic (fcc) lattice provides the densest packing compatible with cubic lattices and supercells built from them, we initialized vacuum WFs (VWFs) on an fcc grid spanning a $3\!\times\! 3\!\times\! 3$ conventional cubic supercell. Going beyond the $d=1,2$ numerical experiments of~\cite{doi:10.1137/18M1167164}, our tests (Fig.~\ref{fig:perturbed_centers}(a–d)) support the conjecture above for $d=3$: VWFs with centers randomly displaced by up to 30\% of the ideal nearest-neighbor spacing consistently relax back to a regular grid, indicating the close-packed arrangement is likely the global minimum for a cubic supercell. As a specific benchmark, applying random perturbations of the center locations by 30\% of this spacing, followed by several hundred localization steps, yields symmetry-equivalent VWFs that share the same spatial variance to high numerical precision, $\langle r^2 \rangle - \langle r \rangle^2 = 2.80055(8)$ bohr\textsuperscript{2}, and have centers at the expected locations to within an rms deviation of $4.16\times 10^{-5}$ bohr. Anticipating use in systems with hexagonal symmetry, we find corresponding results for the hexagonal close-packed lattice, with stability up to 10\% distortion of the initial locations for a $2\!\times\! 2\!\times\! 4$ system (Fig.~\ref{fig:perturbed_centers}(e–f)).

Together, these results unify the canonical and disentangled perspectives and show that optimal localization selects close-packed lattices as stable, symmetry-equivalent solutions for vacuum WFs. The regularity and stability of such bases enable low-cost extension of vacuum regions, providing a natural foundation for scalable, DFT-accurate treatments of vacuum regions without the need to enlarge the underlying first-principles supercell.

\emph{Applications---} To demonstrate the ability of our approach to treat both material-bound wave functions and vacuum scattering states within a unified framework, we apply our vacuum Wannier–function formalism to photoemission from monolayer homopolar graphene (Gr) and heteropolar hexagonal boron nitride (h-BN).

A key observable in the photoelectron distribution, central to many applications, is the mean transverse energy (MTE) of the emitted electrons. For example, the beam brightness of a photocathode source scales as \(B\!\propto\!1/\mathrm{MTE}\) \cite{10.1063/1.5053082}, so that brightness is maximized by minimizing MTE.

Likewise, minimizing the energy spread \(\sigma_{E_\mathrm{L}}\) is critical for ultrafast electron diffraction applications. In particular, any uncorrelated rms spread $\sigma_{E_\mathrm{L}}$ in the longitudinal distribution maps directly into arrival-time jitter via vacuum dispersion and into residual chirp that resists compression; minimizing this spread is therefore essential for achieving the shortest, most stable electron pulses at the sample and for avoiding wavelength-spread blurring in the resulting diffraction~\cite{RevModPhys.94.045004}.

As a baseline for comparison, we note that, for semiconductors and insulators, a simple effective-mass model with a \(k\)-independent optical matrix element yields
\begin{equation}
\sigma_{E_\mathrm{L}} = \frac{2}{3\sqrt{5}}E_{\text{excess}}\approx0.298E_{\text{excess}},
\label{eqn:sigmaEL_estimate}
\end{equation}
which is independent of the effective masses, and
\begin{equation}
\frac{\mathrm{MTE}}{E_\text{excess}}
= \frac{2}{3}\frac{\sum_i \left(\dfrac{m_i^*}{m_e + m_i^*}\right)^{3/2}}
{\sum_i \left(\dfrac{m_i^*}{m_e + m_i^*}\right)^{1/2}},
\label{eqn:mte_estimate}
\end{equation}
where the $m_i^*$ are the effective masses of the valence bands contributing to emission and $m_e$ is the free-electron mass. For a single parabolic band with $m^*=m_e$, Eq.~\eqref{eqn:mte_estimate} reduces to $\mathrm{MTE}/E_{\text{excess}} = 1/3$, the well-known Dowell--Schmerge result~\cite{10.1063/1.4922146,PhysRevSTAB.12.074201}.

\begin{figure}[h!]
    \centering
    \includegraphics[width=1\linewidth]{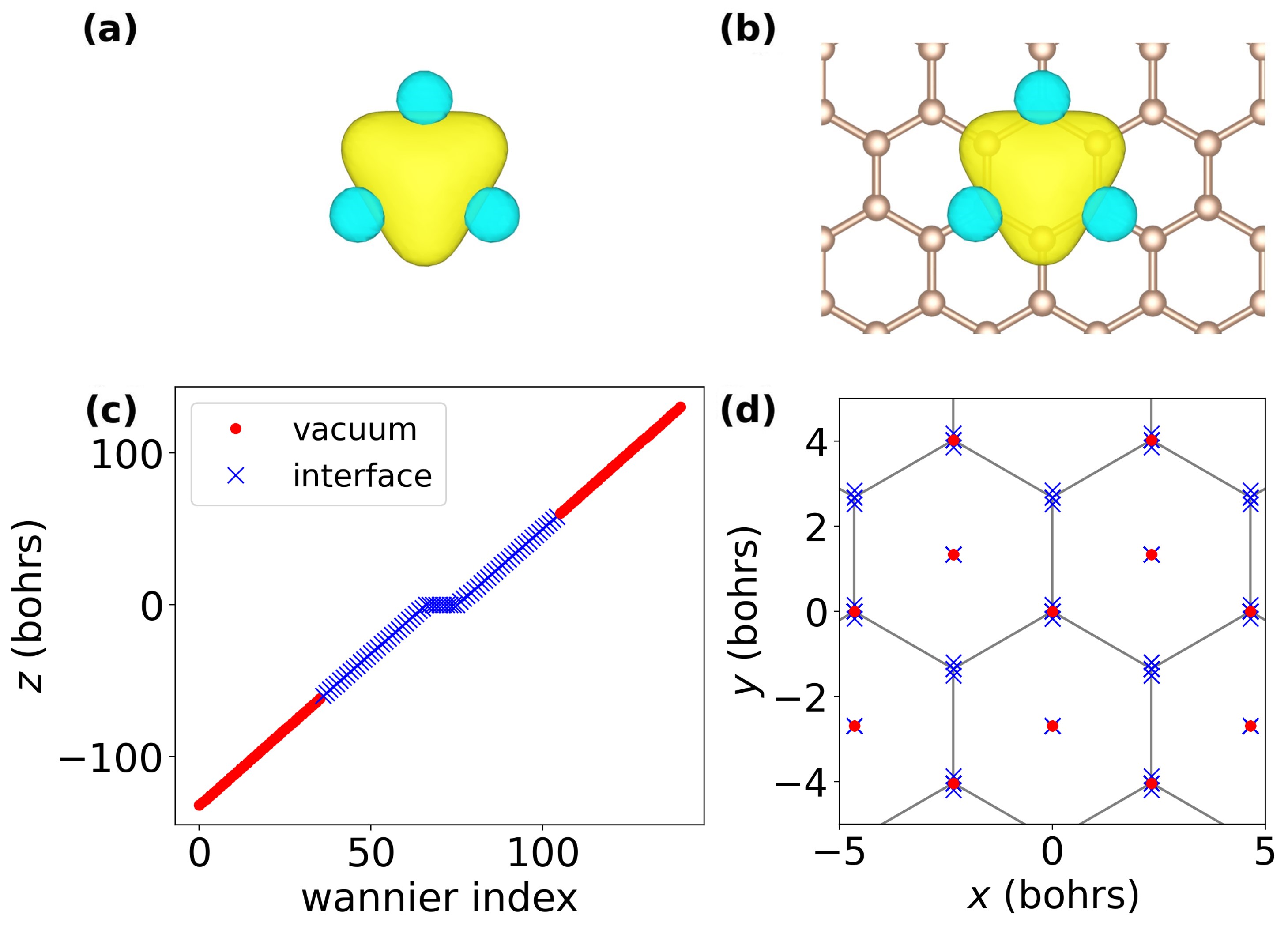}
    \caption{Vacuum Wannier functions (VWFs): (a) disentangled VWF in the empty lattice and (b) numerically optimized VWF located $\sim 50$\,\AA\  above a graphene sheet. The close similarity demonstrates that the analytic theory and realistic slab calculations yield the same compact, regularized functions. Final Wannier-center locations after optimization projected along (c) the normal and (d) transverse planar directions. Yellow and blue surfaces in (a) and (b) represent equal‑but‑opposite sign contour levels; blue $\times$ symbols and red circles in (c) and (d) denote interfacial and deep-vacuum Wannier functions, respectively. The final optimized Wannier functions show striking regularity in location and spacing.}
    \label{fig:gr_isosurfaces}
\end{figure}

To obtain the \emph{ab initio} results below, we seed the Marzari–Vanderbilt disentanglement procedure with initial Gaussian orbitals at regular positions, which are then projected onto the band space, and then disentanglement is allowed to proceed. This prescription ensures that the disentanglement includes the relevant free-electron-like states and biases the gauge toward the desired vacuum Wannier arrangement. Through extensive numerical tests, we find that this approach consistently yields Wannier functions in the deep-vacuum region that take the expected vacuum form (compare Fig.~\ref{fig:gr_isosurfaces}(a,b)) and exhibit nearly exact periodic placement. The latter is evident in the Wannier-center distributions for a graphene slab (Fig.~\ref{fig:gr_isosurfaces}(c,d)). Along the surface normal, the vacuum-associated centers (Fig.~\ref{fig:gr_isosurfaces}(c)) exhibit regular spacing of $2.000(1)$~bohr, with a small shift for the centers closer to the sheet and convergence to a constant spacing deep in vacuum. In the transverse plane, the vacuum-center projections (Fig.~\ref{fig:gr_isosurfaces}(d)) remain tightly clustered at the origin and at the upper-left vertex of the Wigner-Seitz cell to within $10^{-10}$~bohr. The sheet-associated Wannier functions, on the other hand, are centered near the two atoms of the honeycomb basis, with slight deviations due to hybridization with the vacuum Wannier functions.

\begin{figure}[h!]
    \centering
    \includegraphics[width=1\linewidth]{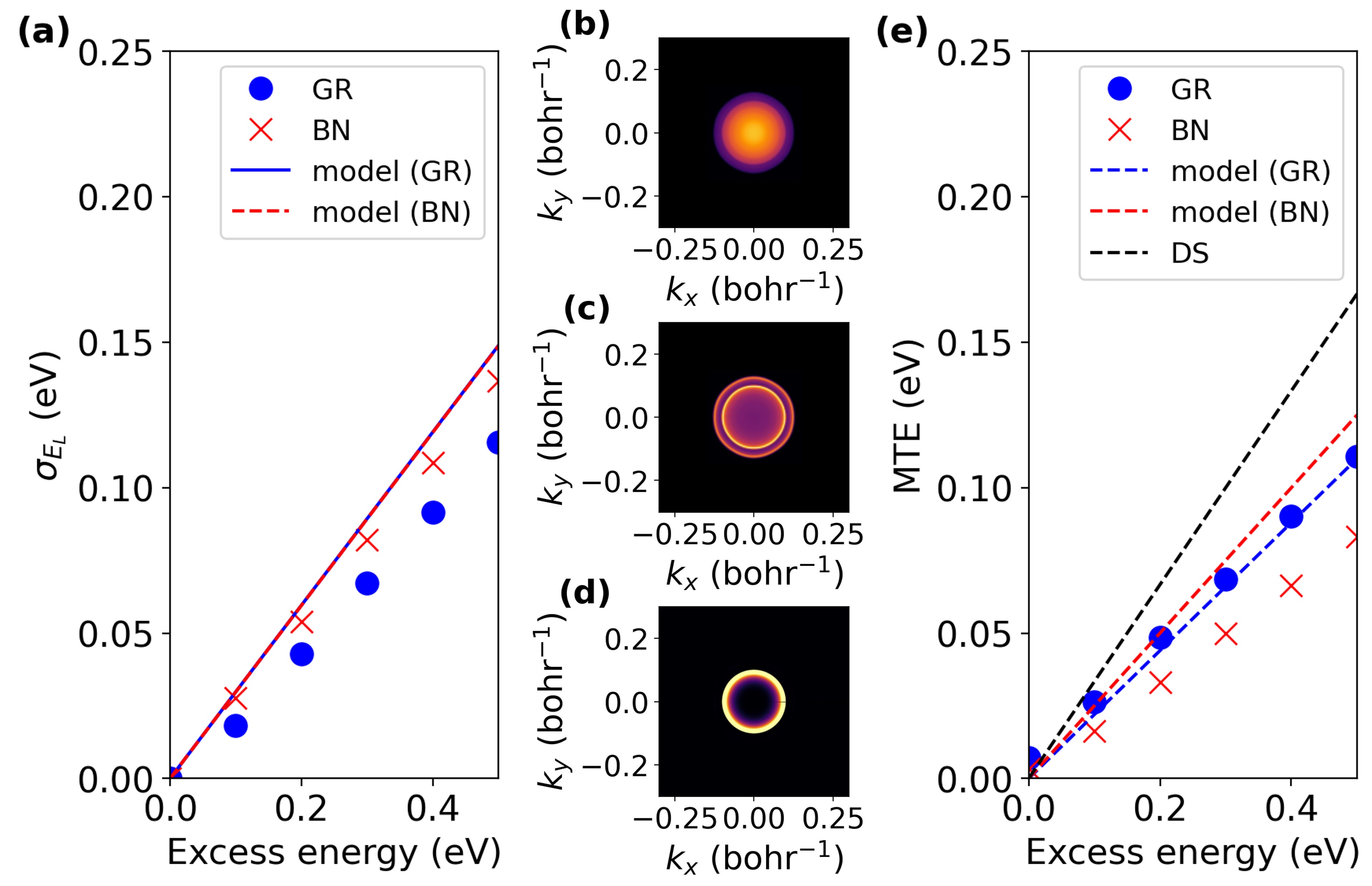}
    \caption{Photoemission results for graphene (Gr) and hexagonal boron-nitride (h-BN). (a) rms widths of the longitudinal energy distributions. (b–d) Transverse momentum distributions for h-BN using (b) the full theory, (c) constant matrix elements, and (d) matrix elements based on plane‑wave final states. (e) Mean transverse energy (MTE) as a function of excess photon energy: monolayer graphene (blue), boron nitride (red), and the Dowell–Schmerge model (black). Explicit calculations are shown as circles (graphene) and $\times$ symbols (h-BN), with the effective‑mass model in Eq.~\eqref{eqn:mte_estimate} shown as dashed lines.}
    \label{fig:TMD results}
\end{figure}

With the vacuum Wannier basis established, we now proceed to quantitative photoemission calculations, beginning with the longitudinal energy spread \(\sigma_{E_\mathrm{L}}\), the key quantity to be minimized for ultrafast electron diffraction (UED). Fig.~\ref{fig:TMD results}(a) shows our simple-model and \emph{ab initio} predictions of the corresponding rms spreads. Linear fits to the \emph{ab initio} data yield \(\sigma_{E_\mathrm{L}}=\alpha\,E_{\text{excess}}\) with \(\alpha_{\mathrm{Gr}}=0.235\) and \(\alpha_{\mathrm{h-BN}}=0.273\) for graphene and h-BN, respectively, with graphene falling noticeably below the simple-model value of 0.298, underscoring the importance of first-principles analysis and identifying graphene as the superior candidate for UED applications.

We next consider the mean transverse energy (MTE). For both Gr and h-BN, photoemission originates from two degenerate valence bands at $\Gamma$ with downward curvature. From DFT, we extract effective masses $0.66\,m_e$ and $0.35\,m_e$ for Gr, and $0.81\,m_e$ and $0.43\,m_e$ for h-BN. Inserting these into Eq.~\eqref{eqn:mte_estimate} yields slope estimates of $0.22$ (Gr) and $0.25$ (h-BN), both notably smaller than the $1/3$ Dowell–Schmerge result. Our \emph{ab initio} results (Fig.~\ref{fig:TMD results}(e)) exhibit the expected linear behavior, with best-fit slopes of $0.208$ (Gr) and $0.160$ (h-BN). In this case, graphene agrees well with the effective-mass estimate, whereas boron nitride deviates substantially.

To understand the origin of this deviation, we examine the transverse-momentum distribution (TMD). Fig.~\ref{fig:TMD results}(b) shows the fully computed TMD for h-BN, whose filled zone center at $k_{\parallel}=0$ concentrates spectral weight at low transverse momentum and thereby reduces the MTE relative to simplified models. By contrast, Fig.~\ref{fig:TMD results}(c) shows the result obtained using constant optical matrix elements, which reduces the calculation to a joint density of states, corresponding to the simplified model discussed above. The resulting pronounced depletion at the zone center reveals the origin of the systematic overestimation of the MTE in simplified models.

Having established the importance of matrix elements, the next question is how they should be computed. The most common approximation is the first Born approximation, in which the outgoing electron is treated as a plane wave. The resulting distribution, shown in Fig.~\ref{fig:TMD results}(d), closely resembles the constant matrix-element result in Fig.~\ref{fig:TMD results}(c), retaining a hollow zone center and therefore producing similarly inflated MTEs. This hollow appearance—specifically the vanishing intensity at $k_{\parallel}=0$—follows directly from the symmetry of the outgoing plane-wave state in the Born approximation. Consequently, including matrix elements alone is insufficient: capturing the correct physics requires going beyond the Born approximation. At higher orders, the broken inversion symmetry of h-BN is imprinted on the final-state wavefunction, permitting finite intensity at the zone center, as realized in our full scattering-based treatment.

By contrast, no analogous anomaly appears in graphene: the fully computed MTE shown in Fig.~\ref{fig:TMD results}(e) closely follows the expectations of the simplest theoretical treatments, with no qualitative change upon including full scattering effects. This absence of unusual behavior can be understood on symmetry grounds. Graphene is inversion symmetric, which enforces vanishing optical matrix elements at the zone center for normally incident light. As a result, its momentum distributions exhibit a robust central depletion that is preserved whether the matrix elements are evaluated within the first Born approximation or the full scattering series is included.

The stark contrast with h-BN highlights the essential role of symmetry breaking: in the absence of inversion symmetry, higher-order scattering processes generate substantial zone-center weight and dramatically suppress the MTE. Capturing this physics requires access to the full Born series over dense momentum grids, which is readily achieved within our Wannier-based formalism and underlies the central theoretical advance of this work.

\emph{Conclusion.---} We have established a first-principles theory of vacuum Wannier functions that resolves optimization pathologies in their numerical construction in deep vacuum regions. Analytic solutions of the canonical and disentangled localization procedures indicate that optimally localized vacuum Wannier functions organize on regular close-packed lattices. This emergent geometric structure guides practical basis construction and enables scalable extension of vacuum regions without enlarging the underlying electronic-structure calculations, thereby providing direct access to dense $k$-space constructions of Born-series scattering states at interfaces. The resulting framework unifies tight-binding and continuum descriptions of emission and scattering, opening a systematic route to predictive treatments of vacuum-coupled phenomena.

\emph{Acknowledgments.---}
This work was supported by the U.S. National Science Foundation under Award PHY-1549132, the Center for Bright Beams.

\medskip
\renewcommand\refname{}
\renewcommand\bibsection{\hrule height 0.4pt\vspace{10pt}}
\bibliographystyle{unsrt}
\bibliography{ref}

\clearpage

\section*{End matter}

\emph{Regular spacing of canonically constructed vacuum Wannier functions.---} 
If we Wannierize a single band in $d=1$ dimensions without disentanglement, the well-known result is the sinc WF.
\begin{equation} \label{eq:sinc}
W(x) \;=\; \sqrt{\tfrac{a}{2\pi}}\int_{-\pi/a}^{\pi/a} dk \; e^{ikx}
= \sqrt{\tfrac{a}{2\pi}} \;\frac{\sin(\pi x/a)}{x},
\end{equation}
whose decay is too slow to give a finite second moment.

In practical \emph{ab initio} calculations we work in three dimensions and typically require multiple Wannier functions to capture higher-energy vacuum scattering states. To define localized WFs in this setting, we consider a large vacuum supercell with Bravais lattice vectors $\{\bm{a}_1, \dots, \bm{a}_d\} \subset \mathbb{R}^d$, which yields a correspondingly small first Brillouin zone. Rather than restricting to a single WF associated with the lowest band of this zone, and thereby including only very low energy plane waves, we include all plane-wave states folded into it from an $\{N_1 \times \dots \times N_d\}$ reconstruction. The resulting basis is equivalent to the lowest-band plane waves of a reduced real-space cell spanned by $\bm{a}_i/N_i$. The corresponding WFs then capture higher energies while remaining naturally centered on the regular lattice defined by these vectors, thereby ensuring the periodicity needed to extend vacuum regions in practical \emph{ab initio} calculations with essentially no additional computational cost.

This construction can be understood formally as the projection of a set of position eigenstates $|\bm{\mathcal{R}}\rangle$ onto the truncated plane-wave subspace, yielding a basis of regularly spaced Wannier functions within the original unit cell. Defining the projector
\begin{equation}
\hat{P}_{\mathcal{B}_{\text{red}}} = \frac{1}{(2\pi)^d} \int_{\mathcal{B}_{\text{red}}} d^d\bm{k}\, |\bm{k}\rangle \langle \bm{k}|
\end{equation}
as an integral over the Brillouin $\mathcal{B}_\text{red}$ zone of the reduced real-space cell, 
the vacuum WF centered at $\bm{\mathcal{R}}$ is then
\begin{equation} \label{eq:WR}
|W(\bm{\mathcal{R}})\rangle = \sqrt{\tfrac{V_0}{N_{\text{tot}}}}\, \hat{P}_{\mathcal{B}_{\text{red}}} |\bm{\mathcal{R}}\rangle,
\end{equation}
where $|\bm{\mathcal{R}}\rangle$ is a position eigenstate. From this, direct substitution will recover Eq.~\eqref{eq:sinc} with $N_1=1$ in one dimension and $\mathcal{R}=0$. For the general case, a straightforward calculation using Eq.~\eqref{eq:WR} and the identity $\bra{\bm{\mathcal{R'}}} [\hat{\bm{r}}, \hat{P}_{\mathcal{B}_{\text{red}}}] \ket{\bm{\mathcal{R}}}
    = -i \int_{\partial \mathcal{B}_{\text{red}}} d^{d-1} k\, \hat{\bm{n}}(\bm{k})\, e^{i \bm{k} \cdot (\bm{\mathcal{R'}} - \bm{\mathcal{R}})}$, then gives
\begin{align}
\langle W(\bm{\mathcal{R}})|\hat{\bm{r}}|W(\bm{\mathcal{R'}})\rangle 
&= \frac{V_0}{N_{\text{tot}}} \langle \bm{\mathcal{R}} | \hat{P}_{\mathcal{B}_{\text{red}}} \hat{\bm{r}} \hat{P}_{\mathcal{B}_{\text{red}}} | \bm{\mathcal{R'}} \rangle \nonumber \\
&= \bm{\mathcal{R'}}\, \delta_{\bm{\mathcal{R}}, \bm{\mathcal{R'}}},
\end{align}
so that this construction of the WFs diagonalizes the projected position operator, meaning the typical gauge-dependent part of the localization functional, $\tilde{\Omega}=\sum_{\mathcal{R}'\ne\mathcal{R}} |\langle W(\bm{\mathcal{R}})|\hat{\bm{r}}|W(\bm{\mathcal{R'}})\rangle|^2$~\cite{RevModPhys.84.1419}, is exactly zero. This gives a set of sinc-like Wannier functions naturally arranged on a regular grid with minimal gauge-dependent spread, even though their second moments diverge.

\emph{Analytic disentanglement and regular spacing.---} To cure the divergence for canonically constructed vacuum Wannier functions, we here employ the standard disentanglement procedure discussed in the text and provide a fully analytic solution in the $d = 1$ case, constructing $N$ Wannier functions from the lowest $N$ free-electron bands.
 (For simplicity of presentation, we take $N$ to be even, but the final results generalize naturally to the odd case.)  In the plane-wave basis normalized to a unit cell of length $a$, we define \(\braket{x}{m G_0}\equiv e^{i m G_0 x}/\sqrt{a}\) with \(G_0=2\pi/a\). The initial projector onto the lowest \(N\) bands at fixed \(k\) then can be written as
\begin{eqnarray*}
P(k)&=&\ket{\operatorname{sgn}(-k)\,\tfrac{N}{2}G_0}\bra{\operatorname{sgn}(-k)\,\tfrac{N}{2}G_0}\\
&&\ \ \ \ \ \ \ \ \ \ \ +\!\!\!\!\!\!\!\sum_{m=-(\tfrac{N}{2}-1)}^{\tfrac{N}{2}-1}\!\!\!\!\!\!\!\ket{mG_0}\bra{mG_0}.
\end{eqnarray*}
Here, \(\operatorname{sgn}(-k)\,\tfrac{N}{2}G_0\) selects the \(N\)th band across the zone, accounting for the switch in the \(\pm \tfrac{N}{2}\) component when \(k\) crosses the zone center (Fig.~\ref{fig:1D bands}(a) for $N=2$ bands). The discontinuity in $\operatorname{sgn}(-k)$ makes \(P(k)\) discontinuous at \(k=0\), generating a Dirac delta in \(\partial_k P(k)\) there. Substituting into \eqref{eq:Omega_I} then yields a formal \(\delta(k)^2\) term and hence a divergent \(\Omega_I\) for this vacuum (empty-lattice) case.

To cure this pathology, we employ the disentanglement procedure described in the text. Using the minimally required outer window (Fig.~\ref{fig:1D bands}(a), upper dashed line), we must construct a single supplemental band formed as a unitary combination of \(\ket{\pm\tfrac{N}{2}G_0}\). A simple choice that ensures the final Wannier functions are real is then
\begin{equation}\label{eq:mix}
|u_N(k)\rangle = \cos \theta(k)\cdot |\tfrac{N}{2}\cdot G_0\rangle + \sin\theta(k)\cdot|-\tfrac{N}{2}\cdot G_0\rangle,
\end{equation}
where $\theta(k)$ is a mixing function with $\theta(k<-k_0)=0$ and $\theta(k>k_0)=\pi/2$, ensuring that $\ket{u_N(k)}=\ket{\mathrm{sgn}(-k)\tfrac{N}{2}\cdot G_0}$ everywhere except near the band-crossing point, which is the source of the divergence in $\Omega_I$ for the entangled case.  Varying $P(k)$ and setting the first variation of $\Omega_I$ to zero gives the stationarity condition for the spread functional in Eq.~\eqref{eq:Omega_I}, 
\begin{equation}
[\partial_k^2 P(k),P(k)]=0,
\end{equation}
which, in the two-state subspace of Eq.~\eqref{eq:mix}, reduces to
\begin{align}
[\partial_k^2 P(k),P(k)] =&\ \theta''(k)\Bigg(\ket{-\frac{N}{2}\cdot G_0}\bra{\frac{N}{2}\cdot G_0}\nonumber\\
&-\ket{\frac{N}{2}\cdot G_0}\bra{-\frac{N}{2}\cdot G_0}\Bigg)=0.
\end{align}
Clearly, this is solved by $\theta''(k)=0$, yielding linear interpolation that satisfies the boundary conditions at $\pm k_0$, 
\begin{equation}
\theta(k)=\frac{\pi}{4k_0}(k+k_0), \quad k\in[-k_0,k_0].
\end{equation}

The resulting disentangled $N$-band subspace, when viewed in an extended-zone scheme, becomes equivalent to a single-band problem in a unit cell of size $a/N$. Upon Fourier transforming the resulting band, the corresponding set of $N$ maximally localized WFs is $W(x-an/N)$ for $n=0,\ldots,N-1$, where
\begin{equation}\label{eqn:exact 1d}
\begin{aligned}
W(x)=\frac{1}{2 \pi}\sqrt{\frac{a}{N}} \Bigg[ &
   \sin\!\big((\frac{N}{2}G_0 - k_0)x\big)\,
   \left(
      \frac{2}{x}
      - \frac{2x}{x^2 - (\tfrac{\pi}{4k_0})^2}
   \right) \\[6pt]
   & - \cos\!\big((\frac{N}{2}G_0 + k_0)x\big)\,
   \left(
      \frac{\tfrac{\pi}{2k_0}}{x^2 - (\tfrac{\pi}{4k_0})^2}
   \right)
\Bigg],
\end{aligned}
\end{equation}
which also holds for odd $N$. In real space, because the asymptotic behavior of this function is $|W(x)|\sim 1/x^2$, the variance is finite and the divergence is cured. Finally, one can prove that these functions achieve the absolute minimum possible gauge-dependent localization, $\tilde\Omega$.

\emph{Proof of orthonormality and maximal localization.---} For completeness, we here provide the proof that our disentangled Wannier functions are both orthonormal and maximally localized in the sense of the Marzari–Vanderbilt spread functional. The derivation proceeds in two steps: (i) establishing orthogonality of Wannier functions translated by integer multiples of the reduced cell spacing $\Delta=a/N$, and (ii) showing that the projected position operator is diagonal in this basis, which guarantees vanishing off-diagonal spread and hence maximal localization.

These demonstrations proceed more naturally in the following Fourier representation:
\begin{equation}
W(x)=\frac{1}{\sqrt{2\pi}}\int \! \mathrm{d}q\, e^{iqx}\,\widetilde W(q),
\end{equation}
\begin{equation}
\widetilde W(q)=\frac{1}{\sqrt{2\pi}}\int \! \mathrm{d}x\, e^{-iqx}\,W(x).
\end{equation}
From the exact 1D formula (Eq.~\eqref{eqn:exact 1d}), the momentum-space Wannier function centered at the origin is then
\begin{equation}
\widetilde{W}(q)
=\sqrt{\frac{a}{2\pi N}}
\begin{cases}
\sin\!\bigl[\dfrac{\pi}{4k_0}\,(q+\frac{N\pi}{a}+k_0)\bigr], & \!\!\!\!\!|q+\frac{N\pi}{a}|<k_0,\\[4pt]
1, & \!\!\!\!\! |q|<\frac{N\pi}{a}-k_0,\\[4pt]
\cos\!\bigl[\dfrac{\pi}{4k_0}\,(q-\frac{N\pi}{a}+k_0)\bigr], & \!\!\!\!\! |q-\frac{N\pi}{a}|<k_0,\\[4pt]
0, & \text{otherwise}
\end{cases}
\end{equation}
where the intervals defined for the first and third cases are referred to as ``shoulders." Here \(N\) is the number of disentangled bands, and the real-space spacing between Wannier centers is
$\Delta \equiv \frac{a}{N}$. A shift by \(\ell\Delta\) in real space corresponds to a phase in momentum space:
\begin{equation}
\widetilde W_j(q)=e^{-iq\,\ell\Delta}\,\widetilde W(q).
\end{equation}

Now, to establish orthonormality, let \(\ket{m}\) and \(\ket{n}\) denote Wannier functions centered at \(x=m\Delta\) and \(x=n\Delta\), respectively. Their overlap evaluates to
\begin{equation}
\begin{aligned}
\braket{m}{n}
=\frac{a}{2\pi N}\Bigg[&
\frac{2\,\sin\!\bigl((\tfrac{N\pi}{a}-k_0)\,\Delta(m-n)\bigr)}{\Delta(m-n)}\\
&+\frac{2(-1)^{m-n}\,\sin\!\bigl(k_0\,\Delta(m-n)\bigr)}{\Delta(m-n)}
\Bigg].
\end{aligned}
\end{equation}
Using the identity $\Delta\frac{N\pi}{a}=\pi$ and the sine angle-sum identity on the first term, one finds \(\braket{m}{n}=0\) for \(m\neq n\), while the limit \((m-n)\to 0\) gives \(\braket{m}{m}=1\). Therefore,
\begin{equation}
\braket{m}{n}=\delta_{mn}.
\end{equation}

Next, we turn to matrix elements of the position operator and localization. Because the Wannier functions are translates of a single \(\ket{0}\),
\begin{equation}
\mel{m}{\hat{x}}{n}=n\Delta\,\braket{m-n}{0}+\mel{m-n}{\hat{x}}{0}.
\end{equation}
By orthonormality, the first term is \(n\Delta\,\delta_{mn}\). For the second term, writing \(\ell\equiv m-n\) and using \(\hat{x}=i\,\partial_q\) in momentum space, gives
\begin{equation}
\mel{\ell}{\hat{x}}{0}
= i\int \! \mathrm{d}q\,e^{iq\,\ell\Delta}\,\widetilde W^*(q)\,\partial_q \widetilde W(q).
\end{equation}
Shifting variables on the ``shoulder'' intervals and combining with the plateau, this becomes
\begin{equation}
\mel{\ell}{\hat{x}}{0}
= \frac{a}{2\pi N}\,e^{-i\,k_0\,\ell\Delta}\,\sin(\ell\pi)\,
\int_{0}^{\pi/2}\! \mathrm{d}\theta\;\sin(2\theta)\,e^{i\,\frac{4k_0}{\pi}\theta},
\end{equation}
a term consisting only of the combined ``shoulder'' integrals. Since \(\sin(\ell\pi)=0\) for all integers \(\ell\), we have \(\mel{\ell}{\hat{x}}{0}=0\), and therefore
\begin{equation}
\mel{m}{\hat{x}}{n}=n\Delta\,\delta_{mn}.
\end{equation}
Thus the projected position operator is diagonal in the Wannier basis with eigenvalues located at the lattice of centers \(\{n\Delta\}\). In 1D this implies vanishing off-diagonal spread within the disentangled subspace, i.e.\ our Wannier functions are indeed \emph{maximally localized}.

\smallskip
\noindent\emph{Regularity note.} The sine/cosine ``shoulders'' ensure \(\widetilde W(q)\) and \(\partial_q \widetilde W(q)\) are continuous and compactly supported on the three intervals above, so no boundary terms arise in the manipulations leading to the \(\sin(\ell\pi)\) factor.

\emph{Computational details.---} All plane-wave density functional theory (DFT) calculations were carried out using the JDFTx package~\cite{SUNDARARAMAN2017278}, employing the SG15 Optimized Norm-Conserving Vanderbilt (ONCV) pseudopotentials~\cite{SCHLIPF201536} with a kinetic energy cutoff of 30 Hartree. To ensure proper alignment of the vacuum level, we used the truncated Coulomb interaction for slab geometries~\cite{PhysRevB.73.233103}.

\end{document}